\documentclass[twoside]{dis07}
\usepackage[latin1]{inputenc}
\usepackage[dvips]{graphicx,epsfig,color}
\usepackage{wrapfig,rotating}
\usepackage{amssymb,amsmath,array}

\def\msbar{\ensuremath{{\rm{\overline{MS}}}}}

\newcommand{\gev    }{\ensuremath{\mathrm{GeV}}}

\def\alt{\:\raisebox{-0.5ex}{$\stackrel{\textstyle<}{\sim}$}\:}

\begin{document}
\title{RECENT DEVELOPMENTS IN HEAVY FLAVOUR PRODUCTION
}

\author{G.~KRAMER 
\vspace{.3cm}\\
Universit\"at Hamburg -
II. Institut f\"ur Theoretische Physik \\ 
Luruper Chaussee 149 -
22761 Hamburg - Germany
}


\maketitle
                      
\begin{abstract}
We review one-particle inclusive production of 
heavy-flavoured hadrons in a framework
which resums the large collinear logarithms through the evolution 
of the FFs and PDFs and retains the full dependence on the heavy-quark mass 
without additional theoretical assumptions.
We focus on presenting results for the inclusive cross section for the
production of charmed mesons in $p\bar{p}$ collisions and the comparison
with CDF data from the Tevatron as well as on inclusive $B$-meson production
and comparison with recent CDF data. The third topic is the production of 
$D^\star$ mesons in photoproduction and comparison with recent H1 data from 
HERA.
\end{abstract}

\section{Introduction}
One-particle inclusive production processes provide 
extensive tests of perturbative quantum chromodynamics (pQCD).
In contrast to fully inclusive processes, it is possible to study
distributions in the momentum of the final-state particle and 
to apply kinematical cuts close to the experimental situation.
On the other hand, contrary to even more exclusive cases,
QCD factorisation theorems \cite{Collins:1989gx,Collins:1987pm}
still hold stating that this
class of observables can be computed as convolutions of
{\em universal} parton distribution functions (PDFs) and
fragmentation functions (FFs) with perturbatively 
calculable hard scattering cross sections.
As is well-known, it is due to the factorisation property that
the parton model of QCD has predictive power.
Hence, tests of the universality of the PDFs and FFs are of
crucial importance for validating this QCD framework.
At the same time, lowest-order expressions for the hard scattering
cross sections are often not sufficient for meaningful tests 
and the use of higher order computations is needed.

The perturbative analysis is becoming more involved and interesting
if the observed final state hadron contains a heavy (charm or bottom)
quark. In this case, the heavy-quark mass $m$ enters as an 
additional scale.
Clearly, the conventional massless formalism, also known
as zero-mass variable-flavour-number scheme (ZM-VFNS), can also be
applied to this case, provided the hard scale $Q$ of the process
is much bigger than the heavy-quark mass 
so that terms $m/Q$ are negligible.
However, at present collider energies, most of the experimental data
lie in the kinematic region $Q \gtrsim m$ and it is necessary
to take the power-like mass terms into account in a consistent
framework.

The conventional calculational scheme is the so-called massive scheme or 
fixed-flavour-number scheme (FFNS) \cite{Nason}, in which the 
number of active flavours
in the initial state is limited to $n_f=3$ ($n_f=4$) in the case of massive
charm (bottom) production, and the $c$ ($b$) quark appears only in the final 
state. In this case, the $c$ ($b$) quark is always treated as a heavy 
particle, not as a parton. The actual mass parameter $m$ is explicitly taken
into account along with $p_T$. In this scheme, $m$ acts as a cutoff for the 
initial- and final sate collinear singularities and sets the scale for the
perturbative calculations. A factorisation of these would-be initial- and final
state collinear singularities is not necessary, neither is the introduction of
a FF for the transition $b \rightarrow B$. However at NLO, terms proportional
to $\alpha_s\ln(p_T^2/m^2)$, where $\alpha_s$ is the strong coupling constant, 
arise from collinear gluon emissions by $c$ ($b$) quarks or from branchings
of gluons into collinear $c\bar{c}$ ($b\bar{b}$) pairs. These terms are of
order $O$(1) for large $p_T$ and spoil the convergence of the perturbation 
series. The FFNS with $n_f=3$ ($n_f=4$) should be limited to a rather limited
range of $p_T$, from $p_T=0$ to $p_T \gtrsim m$. The advantage of this scheme,
however, is that the $m^2/p_T^2$ power terms are fully taken into account.

The ZM-VFNS and FFNS are valid in complementary regions of $p_T$, and it is
desirable to combine them in a unified approach that incorporates the 
advantages of both schemes, i.e. to resum the large logarithms, retain the full
finite-$m$ effects, and preserve the universality of the FFs. 
An earlier approach to implement such an interpolation is the so-called 
fixed-order-next-to-leading logarithm (FONLL) scheme, in which the conventional
cross section in the FFNS is linearly combined wit a suitably modified cross 
section in the ZM-VFNS with perturbative FFs, using a $p_T$-dependent weight
function \cite{CGN}. Then the FONLL cross section is convoluted with a 
non-pertubative FF for the $b \rightarrow B$ transition.
  
The subject of this review is the theoretical description of 
one-particle inclusive production of heavy-flavoured hadrons 
$X_h=D,B,\Lambda_c,\ldots$ in a massive variable-flavour-number scheme 
(GM-VFNS). In such a scheme the large collinear logarithms of the 
heavy-quark mass $\ln \mu/m$ are subtracted from the hard scattering cross 
sections and resummed through the evolution of the FFs and PDFs.
At the same time, finite non-logarithmic mass terms $m/Q$
are retained in the hard part and fully taken into account.

In order to test the pQCD formalism, in particular the
universality of the FFs, it is important to provide a 
description of all relevant processes in a coherent framework.
Therefore, it is important to work out the GM-VFNS at
next-to-leading order (NLO) of QCD for all the relevant processes.
Previously, the GM-VFNS has been applied to the following
processes:
$\gamma + \gamma \to D^{\star +} + X$ (direct part) \cite{Kramer:2001gd},
$\gamma + \gamma \to D^{\star +} + X$ (single resolved part) 
\cite{Kramer:2003cw},
$\gamma + p \to D^{\star +} + X$ (direct part) 
\cite{Kramer:2003jw},
$p + \bar{p} \to (D^0,D^{\star +},D^+,D_s^+) + X$
\cite{Kniehl:2004fy,Kniehl:2005mk,Kniehl:2005ej},
where the latter results for hadron--hadron collisions 
also constitute the resolved contribution
to the photoproduction process $\gamma + p \to X_h + X$.

In this contribution, I will review the progress achieved in describing the 
production of heavy-flavoured hadrons $X_h$ in hadron--hadron and 
photon--proton collisions in the GM-VFN scheme as it has been worked out 
recently.  
The main focus will be on the comparison with experimental data from CDF at the
Tevatron for $p+\bar{p} \rightarrow (D^0,D^{*+},D^{+},D^{+}_s)+X$ and
$p+\bar{p} \rightarrow B^{+}+X$ and from H1 at HERA for $\gamma + p \rightarrow
D^{*+} + X$.

\section{Theoretical Framework}
\subsection{GM-VFNS}
The differential cross sections for inclusive heavy-flavoured hadron
production can be computed in the GM-VFNS according to the
familiar factorisation formulae, 
however, with heavy-quark mass terms included in the hard scattering
cross sections \cite{Collins:1998rz}.
Generically, the physical cross sections are expressed as convolutions
of PDFs for the incoming hadron(s), hard scattering cross sections, and
FFs for the fragmentation of the outgoing partons into the observed
hadron. All possible partonic subprocesses are taken into account.
The massive hard scattering cross sections are constructed in a way
that in the limit $m \to 0$ the conventional ZM-VFNS is recovered.
A more detailed discussion of the GM-VFNS and the construction of
the massive hard scattering cross sections can be found
in Refs.\ \cite{Kniehl:2004fy,Kniehl:2005mk}
and the conference proceedings 
\cite{Schienbein:2003et,Schienbein:2004ah,Kniehl:2005st,Baines:2006uw}.
\subsection{Fragmentation Functions}
A crucial ingredient entering these calculation are the 
non-perturbative FFs for the transition
of the final state parton into the observed hadron $X_h$.
For charm-flavoured mesons, $X_c$, such sets of FFs have been
constructed quite some time ago. 
For $X_c=D^{*+}$, FFs were extracted at LO and NLO in the
$\overline{\rm MS}$ factorisation scheme with $n_f=5$ massless quark flavours
\cite{Binnewies:1998xq} from the scaled-energy ($x$) distribution
$d\sigma/dx$ of the cross section of $e^+e^-\to D^{*+}+X$ measured by the
ALEPH \cite{Barate:1999bg} and OPAL \cite{Ackerstaff:1997ki} 
collaborations at CERN LEP1.
%
Recently, this analysis was extended \cite{Kniehl:2005de}
to include $X_c=D^0,D^+,D_s^+,\Lambda_c^+$ by exploiting
appropriate OPAL data \cite{Alexander:1996wy}.
%
In Refs.~\cite{Binnewies:1998xq,Kniehl:2005de}, the starting scales 
$\mu_0$ for the DGLAP evolution of the $a\to X_c$ FFs in the 
factorisation scale $\mu_F^\prime$ have been
taken to be $\mu_0=2 m_c$ for
$a=g,u,\overline{u},d,\overline{d},s,\overline{s},c,\overline{c}$ and
$\mu_0=2 m_b$ for $a=b,\overline{b}$.
The FFs for $a=g,u,\overline{u},d,\overline{d},s,\overline{s}$ were assumed to
be zero at $\mu_F^\prime=\mu_0$ and were generated through the DGLAP evolution
to larger values of $\mu_F^\prime$. 
For consistency with the $\msbar$ prescription for PDFs, we repeated the fits 
of the $X_c$ FFs for the choice $\mu_0 = m_c,m_b$ \cite{Kniehl:2006mw}.
\begin{figure*}[t]
\begin{center}
\begin{tabular}{ll}
{\parbox{6.0cm}{
\epsfig{file=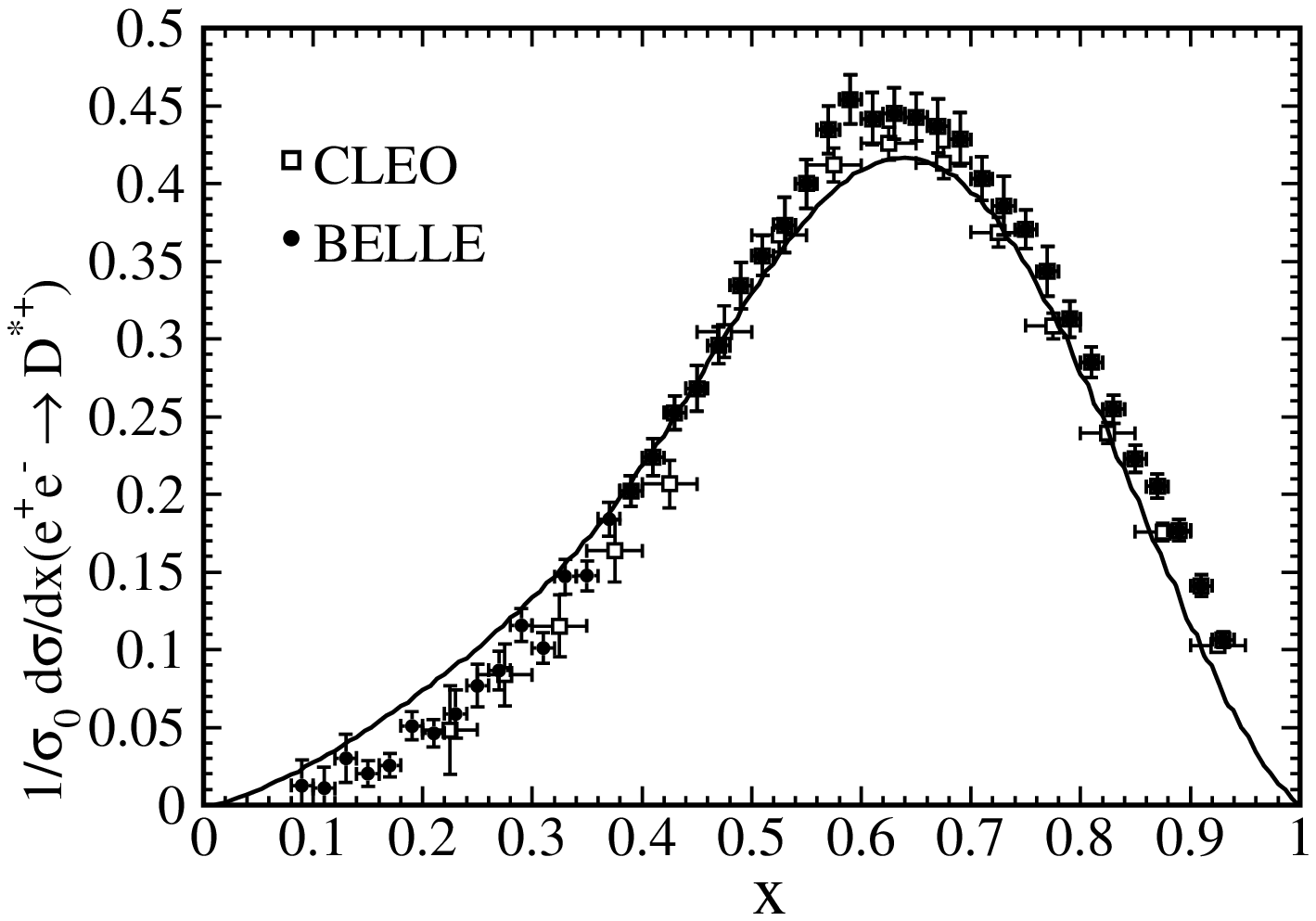,width=5.5cm}
}} &
{\parbox{6.0cm}{
\hspace*{-1.5cm}
\epsfig{file=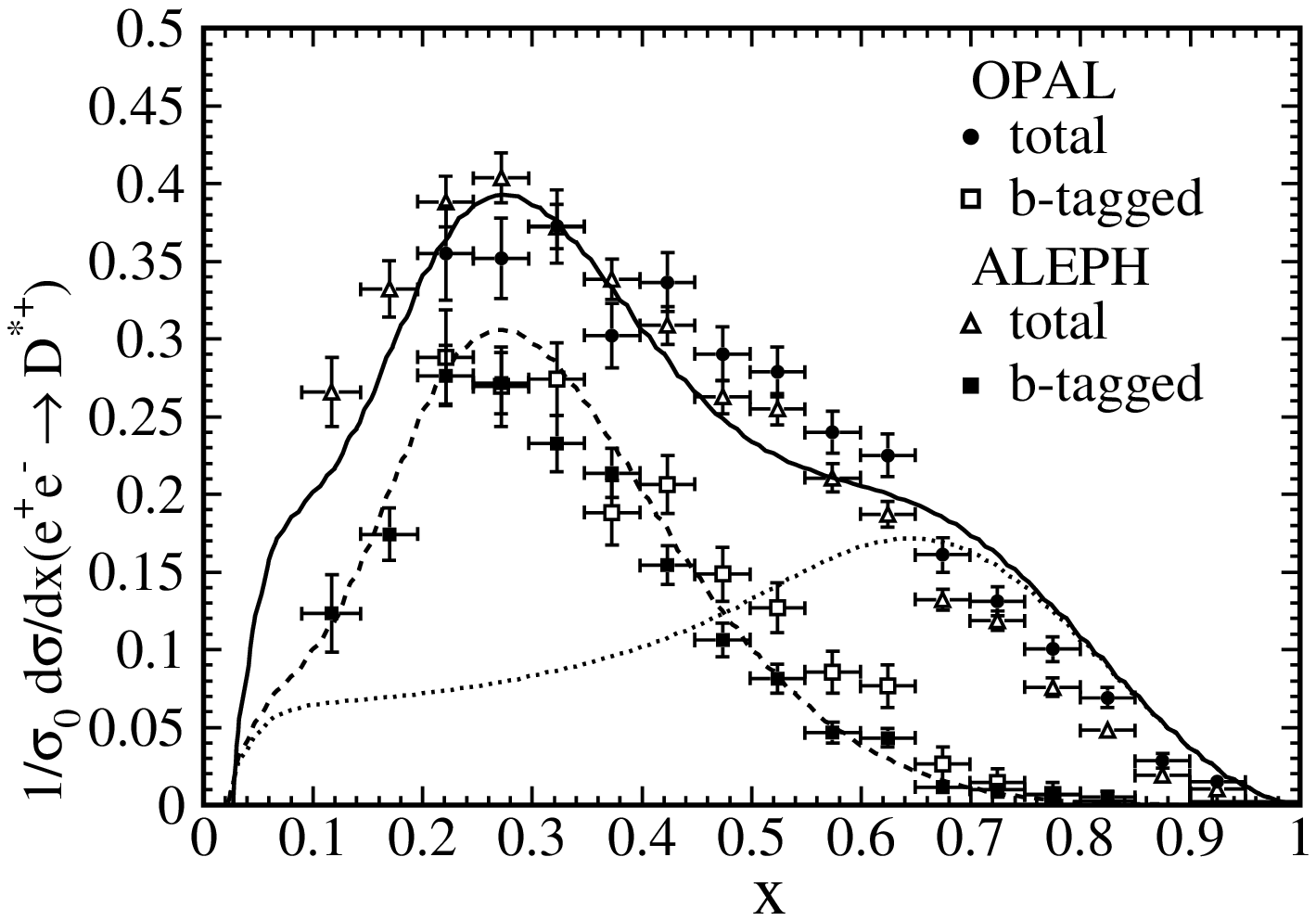,width=5.5cm}
}}
\end{tabular}
\end{center}
\caption{
The cross section for inclusive $D^{*\pm}$ production in $e^+e^-$ annihilation 
evaluated in NLO is compared with from CLEO \protect\cite{Artuso:2004pj} and 
BELLE \protect\cite{Seuster:2005tr} (left) as well as from the ALEPH 
\protect\cite{Barate:1999bg} and OPAL \protect\cite{Ackerstaff:1997ki} data 
(right). The three curves in the right figure correspond  to the 
$Z \rightarrow c\bar{c}$, $Z \rightarrow b\bar{b}$ and full samples.}
\label{fig:fig1}
\end{figure*}
This changes the $c$-quark FFs only marginally, but has an 
appreciable effect on the gluon FF, which is
important at Tevatron energies, as was
found for $D^{*+}$ production in Ref.~\cite{Kniehl:2004fy}.
In the meantime much more accurate data for the inclusive production of
$D^0, D^{+}$ and $D^{*+}$ mesons in $e^+e^-$ annihilation have been
published by the CLEO \cite{Artuso:2004pj} and the BELLE \cite{Seuster:2005tr}
collaborations. With these data new FFs have been constructed. These fits 
were done in the framework of the GM-VFNS, where the
finite charm- and bottom-quark masses were kept in the hard scattering cross
sections. A global fit with the data from CLEO and BELLE at 
10.52 GeV together with the ALEPH \cite{Barate:1999bg} and OPAL 
\cite{Ackerstaff:1997ki} data at the $Z$-resonance are shown in Fig. 1 
\cite{Kneesch:2007}.

\begin{figure*}[t]
\begin{center}
\epsfig{file=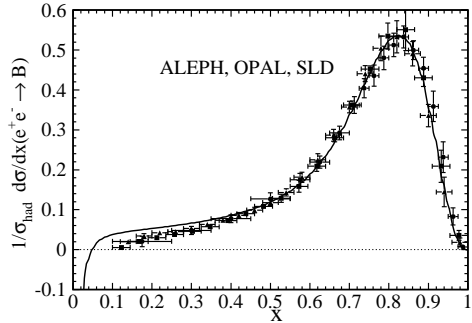,width=7.0cm}
\end{center}
\caption{Comparison of the ALEPH \protect\cite{Heister:2001jg} (circles), 
SLD \protect\cite{Abe:2002iq} (triangles) and OPAL 
\protect\cite{Abbiendi:2003vt} (squares) data with the NLO fits using the power
ansatz. The initial factorisation scale for all partons is $\mu_0$=4.5 GeV.}
\label{fig:fig2}
\end{figure*}

Already many years ago we made an analysis towards FFs for bottom-flavoured
mesons $B^{\pm}$ \cite{Binnewies:1998vm} by using data from the OPAL 
collaboration at LEP1 \cite{Alexander:1995aj}. In the last years much more 
accurate measurements of the inclusive $B$ meson production at the 
$Z$-resonance have been done by
the ALEPH \cite{Heister:2001jg}, SLD \cite{Abe:2002iq} and the OPAL 
\cite{Abbiendi:2003vt} collaborations. With the data in these references 
combined we have performed a new fit to obtain the FFs for 
$q, g, b \rightarrow B$, where, in this case, $q$ are the light quarks 
including $c$. To be consistent with the starting scale of
the PDFs the FFs of these light partons are assumed to vanish at the starting 
scale $\mu_0=m_b=4.5$ GeV and only the $b \rightarrow B$ FF is parametrised 
by the usual power ansatz at the starting scale $\mu_0$. The FFs of the 
light quarks and the gluon are generated via DGLAP evolution at higher scales.
The result of the combined fit is seen in Fig. 2. All three data sets are
consistent with each other and the fit describes the data quite well in the
whole $x$ range, except possibly at rather small $x$ \cite{Kniehl:2007yu}.

\subsection{Input Parameters}
\label{sec:parameters}
For the numerical results presented below we have chosen the following 
input.
For the proton PDFs we have employed the CTEQ6.1M PDFs from the CTEQ 
collaboration \cite{Pumplin:2002vw,Stump:2003yu} and for the charmed meson
fragmentation functions the sets from \cite{Kniehl:2006mw}.
We have set $m_c = 1.5\ \gev$, $m_b = 5\ \gev$ (in the case of charmed meson
production), $m_b = 4.5\ \gev$ (in the case of $B$-meson production) and have 
used the two-loop formula for $\alpha_s^{(n_f)}(\mu_R)$ in the
$\msbar\ $ scheme with $\alpha_s^{(5)}(m_Z) = 0.118$.
The theoretical predictions depend on three scales,
the renormalisation scale $\mu_R$, and the initial- and final-state
factorisation scales $\mu_F$ and $\mu_F^\prime$, respectively.
Our default choice for hadro- and photoproduction has been
$\mu_R = \mu_F = \mu_F^\prime = m_T$, where $m_T = \sqrt{p_T^2 + m^2}$
is the transverse mass. Scale changes are controlled by $\xi_R$ and $\xi_F$,
where $\xi_R=\mu_R/m_T$, $\xi_F=\mu_F/m_T$ and $\xi_F^\prime=\mu_F^\prime/m_T$.

\section{Hadroproduction}
A few years ago the CDF collaboration has published first cross section data
for the inclusive  production of $D^0$, $D^+$, $D^{*+}$, and $D_s^+$ mesons
in $p\bar{p}$ collisions \cite{Acosta:2003ax} obtained in Run II at
the Tevatron at center-of-mass energies of $\sqrt{S} = 1.96$ TeV.
The data come as distributions $d\sigma/dp_T$ with $y$ integrated over 
the range $|y|\le1$ and the particle and antiparticle contributions are
averaged.
\begin{figure*}[t]
\begin{center}
\begin{tabular}{ll}
{\parbox{6.0cm}{
\epsfig{file=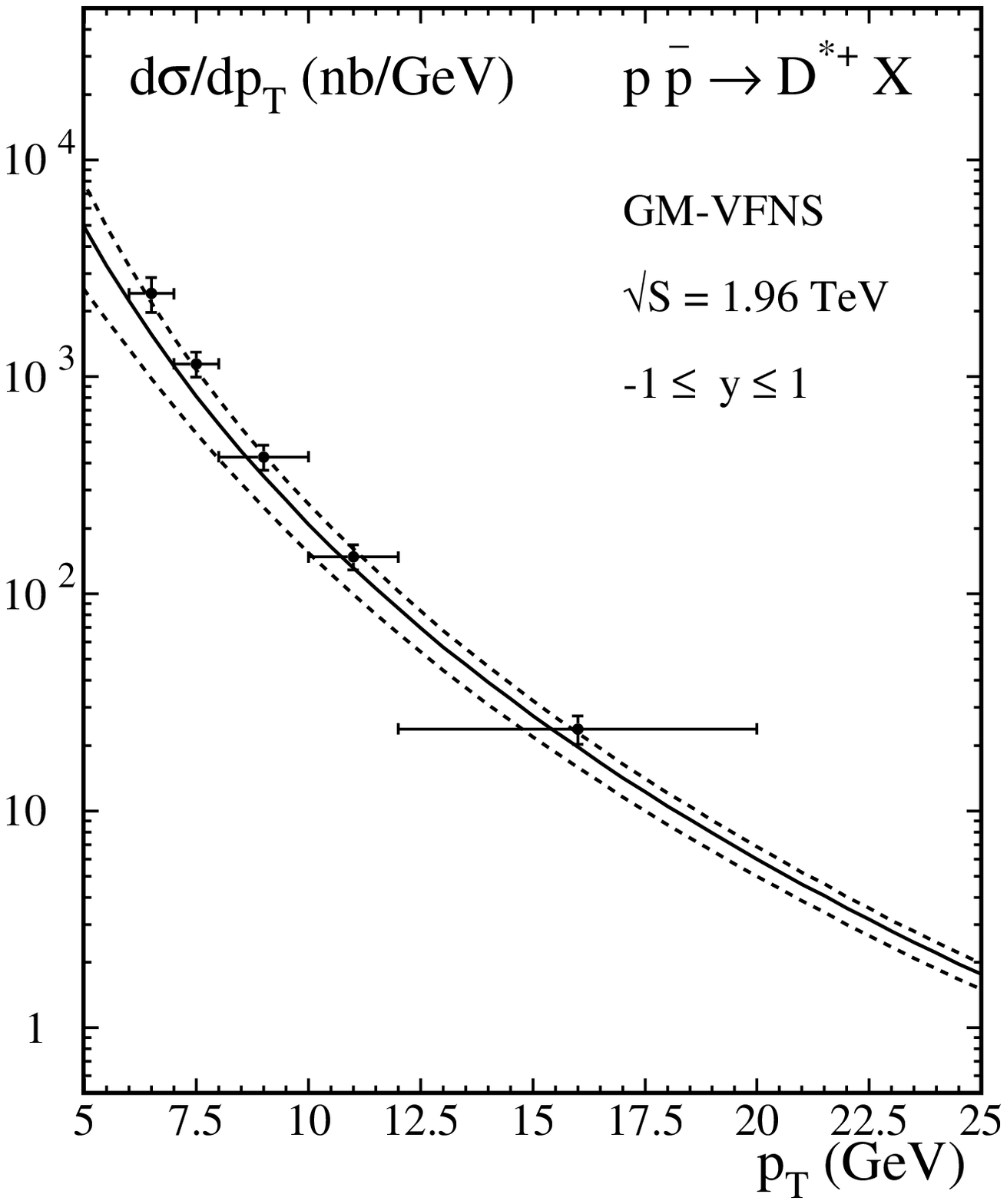,width=6.0cm}
}} &
{\parbox{6.0cm}{
\hspace*{-1.5cm}
\epsfig{file=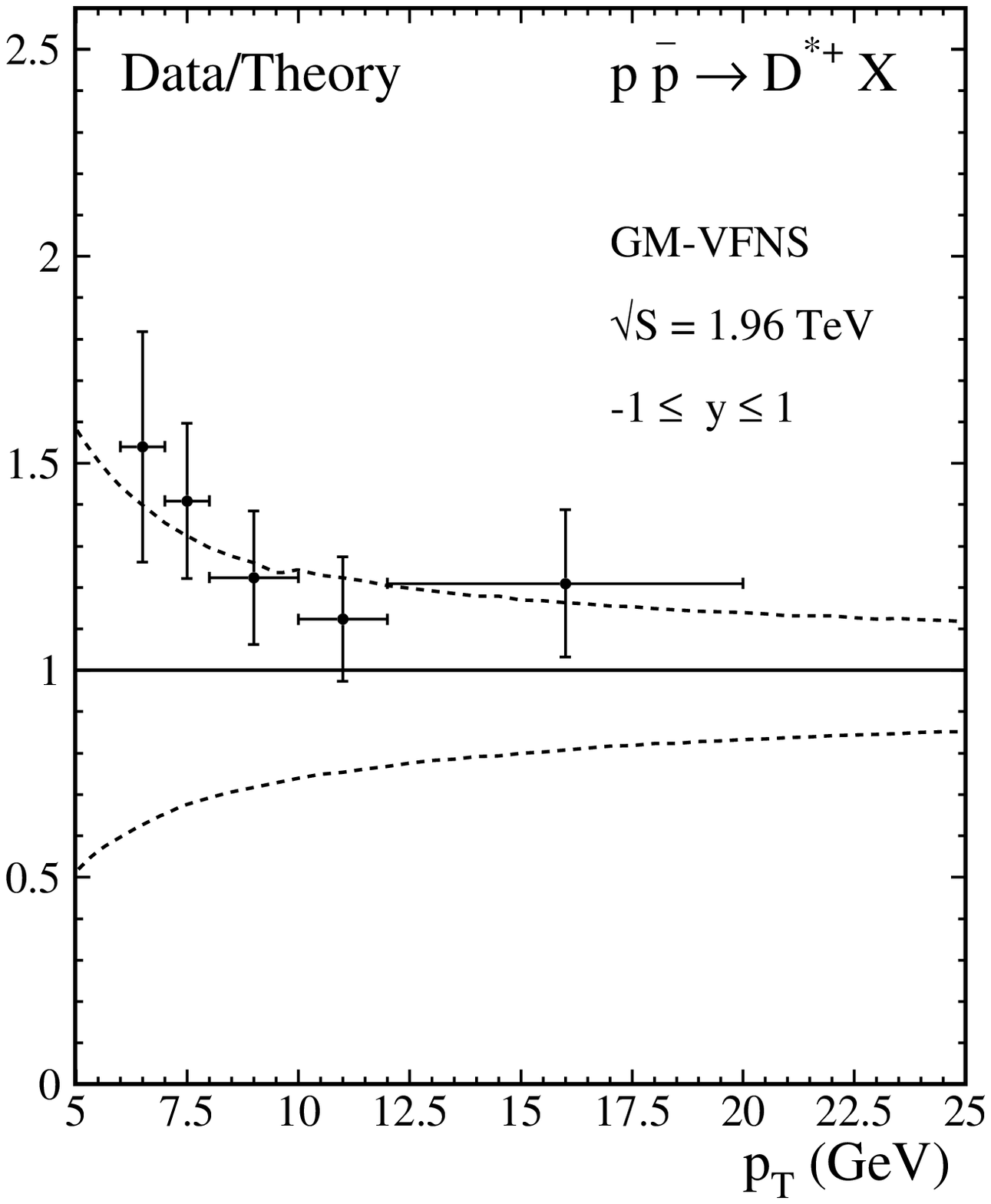,width=6.0cm}
}}
\end{tabular}
\end{center}
\caption{
Comparison of the CDF data \protect\cite{Acosta:2003ax} 
with our NLO predictions for
$D^{*+}$.
The solid line represents our default prediction
obtained with $\mu_R = \mu_F = \mu_F^\prime = m_T$,
while the dashed lines
indicate the scale uncertainty 
estimated by varying $\mu_R$, $\mu_F$, and $\mu_F^\prime$ 
independently within a factor of 2 up and down relative to 
the central values.
The right figure shows
the data-over-theory representation 
with respect to our default prediction.
}
\label{fig:fig3}
\end{figure*}

Our theoretical predictions in the GM-VFNS 
are compared with the CDF data for $D^\star$ mesons on an absolute
scale in Fig.~\ref{fig:fig3} (left) and in the data-over-theory representation
with respect to our default results in Fig.~\ref{fig:fig3} (right).
We find good agreement in the sense that the theoretical and
experimental errors overlap, where
the experimental results are gathered
on the upper side of the theoretical error band, corresponding to a small
value of $\mu_R$ and large values of $\mu_F$ and $\mu_F^\prime$, the $\mu_R$
dependence being dominant in the upper $p_T$ range.
As is evident from Fig.~\ref{fig:fig3} (right), the central data
points tend to overshoot the central QCD prediction by a factor of about 1.5
at the lower end of the considered $p_T$ range, where the errors are largest,
however.
This factor is rapidly approaching unity as the value of $p_T$ is increased.
The tendency of measurements of inclusive hadroproduction in Tevatron run~II
to prefer smaller renormalisation scales is familiar from single jets, which
actually favour $\mu_R=p_T/2$~\cite{field05}. 
It will be interesting to compare
these data with predictions using the most recently constructed fragmentation 
functions based on the BELLE and CLEO data shown above.
For more details and a comparison 
with the data for the $D^0$, $D^+$, and $D_s^+$ mesons
we refer to Ref.\ \cite{Kniehl:2005ej}.

In the GM-VFNS framework we have also calculated the cross
section distribution $d\sigma/dp_T$ of $B$-meson hadroproduction. The 
calculations proceed analogously to the case of $D$ mesons outlined in Ref.
\cite{Kniehl:2004fy}. Now the heavy quark mass $m$ is the $b$ quark mass $m_b$.
The $c$ quark belongs to the group of light quarks $q$, whose mass is put to 
zero. 

The NLO cross section consists of three classes of contributions.\\
Class (i) contains all the partonic subprocesses  with a $b,\bar{b} \rightarrow
B$ transition in the final state that have only light partons
($g,q,\bar{q}$) in the initial state, the possible pairings being $gg$, $gq$,
$g\bar{q}$, and $q\bar{q}$.\\ 
Class (ii) contains all the partonic subprocesses 
with $b,\bar{b} \rightarrow B$ transitions in the final state that also have
$b$ or $\bar{b}$ quarks in the initial state, the possible pairings being $gb$,
$g\bar{b}$, $qb$, $q\bar{b}$, $\bar{q}b$, $\bar{q}\bar{b}$ and 
$\bar{b}\bar{b}$.\\
Class (iii) contains all the partonic subprocesses with a $g,q,\bar{q} 
\rightarrow B$ transition in the final state.\\
In the FFNS only the contributions of class (i) are included, but the full $m$
dependence is retained. On the other hand, in the ZM-VFNS, the contributions
of all the three classes are taken into account, but they are evaluated
for $m=0$. In the GM-VFNS, the class-(i) contribution of the FFNS is matched
to the $\overline{\mathrm{MS}}$ scheme through appropriate subtractions of 
would-be collinear singularities, and is then combined with the class-(ii) and 
class-(iii) contributions of the ZM-VFNS. Thus only the hard scattering cross
sections of class (i) carry explicit $m$ dependence. Specifically, the 
subtractions affect initial states involving $g \rightarrow b\bar{b}$ 
splittings and final states involving $g \rightarrow b\bar{b}$, 
$b \rightarrow gb$ and $\bar{b} \rightarrow g\bar{b}$ splittings, and they
introduce logarithmic dependences on the initial- and final-state factorisation
scales in the hard-scattering cross sections of class (i), which are 
compensated through NLO by the respective factorisation scale dependences by 
the $b$-quark PDF and the $b \rightarrow B$ FF, respectively. It turns out
that the $q$-quark fragmentation contribution is negligible. However, the gluon
fragmentation reaches approximately $50\%$ at small values of $p_T$, and 
somewhat less towards larger values of $p_T$.\\
\begin{figure*}[t]
\begin{center}
\epsfig{file=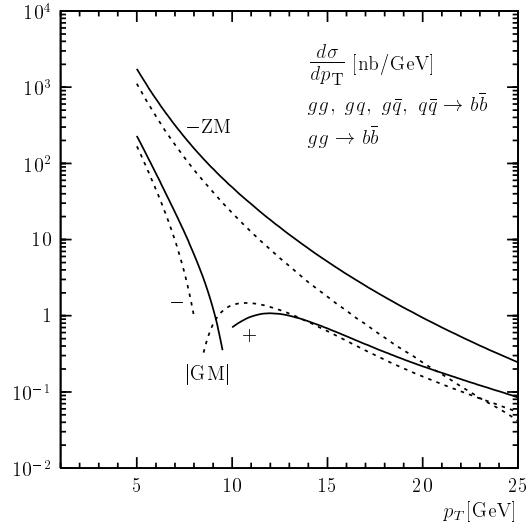,width=7.0cm}
\end{center}
\caption{\small Transverse-momentum distribution $d\sigma/dp_T$ of
$p\bar{p} \rightarrow B +X$ at c.m.\ energy $\sqrt{S}=1.96$~TeV integrated
over the rapidity range $|y| < 1$.
The contributions of class~(i) (solid lines) and their $gg$-initiated parts
(dashed lines) evaluated at NLO in the ZM-VFNS (upper lines) and the GM-VFNS
(lower lines) are compared.}
\label{fig:fig4}
\end{figure*}
The explicit contributions to the hard scattering  cross sections of class (i)
as they contribute to the final result, after all the subtractions are made,
are shown in Fig.~\ref{fig:fig4}.
The results for $m=0$ and finite $m$ are shown in this figure as the upper and 
lower solid lines, respectively.
They constitute parts of the final ZM-VFNS and GM-VFNS results.
In both cases, the contributions of classes (ii) and (iii) for $m=0$ still
must be added to obtain the full predictions to be compared with experimental
data.
The class-(i) contributions in the ZM-VFNS and GM-VFNS schemes are, therefore,
entitled to be negative and they indeed are, for $p_T \alt 76$~GeV and
$p_T \alt 10$~GeV, respectively, as may be seen from Fig.~\ref{fig:fig4}.
Comparing the ZM-VFNS and GM-VFNS results, we notice that the finite-$m$
effects are significant for $p_T \alt 10$~GeV and even cause a sign change for
10~GeV${}\alt p_T \alt 76$~GeV.
However, as will become apparent below, the contributions of class (i) are
overwhelmed by those of classes (ii) and (iii), so that the finite-$m$ effects
are washed out in the final predictions, except for very small values of
$p_T$. 
It is instructive to study the relative importance of the $gg$-initiated
contributions.
They are also included in Fig.~\ref{fig:fig4} for $m=0$ and finite $m$ as the
upper and lower dashed lines, respectively.
They exhibit a similar pattern as the full class-(i) contributions and
dominate the latter in the small-$p_T$ range.
Comparing Fig.~\ref{fig:fig4} with Fig.~2(c) in Ref.~\cite{Kniehl:2004fy}, we 
observe that the relative influence of the finite-$m$ effects is much smaller 
in the $c$-quark case, as expected because the $c$ quark is much lighter than 
the $b$ quark. One can also see from Fig.~\ref{fig:fig4} that the difference 
of the class (i) contributions in the GM-VFNS and ZM-VFNS decrease with 
increasing $p_T$.\\ 
\begin{figure*}[t]
\begin{center}
\epsfig{file=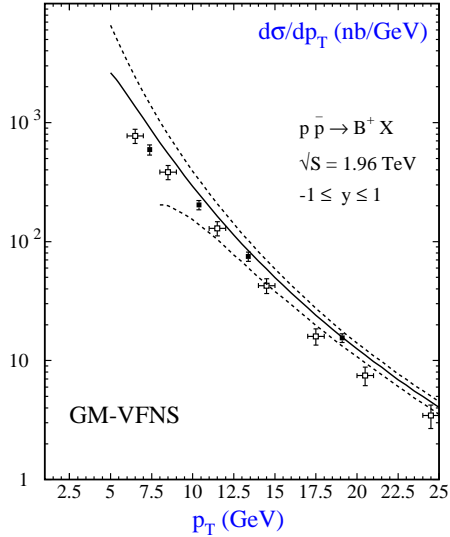,width=7.0cm}
\end{center}
\caption{\small Transverse-momentum distribution $d\sigma/dp_T$ of
$p\bar{p} \rightarrow B +X$ at c.m.\ energy $\sqrt{S}=1.96$~TeV integrated
over the rapidity range $|y| < 1$.
The central NLO prediction with $\xi_R=\xi_F=1$ (solid line) of the GM-VFNS is
compared with CDF data from Refs.~\cite{Acosta:2004yw} (open squares) and 
\cite{Abulencia:2006ps} (solid squares).
The maximum and minimum values obtained by independently varying $\xi_R$ and
$\xi_F$ in the range $1/2\le\xi_R,\xi_F\le2$ with the constraint that
$1/2\le\xi_R/\xi_F\le2$ are also indicated (dashed lines).}
\label{fig:fig5}
\end{figure*}
 
In Fig.~\ref{fig:fig5} we show the comparison of the final prediction,
in which all contributions of classes (i), (ii) and (iii) are combined 
\cite{Kniehl:2007yu} with recent Tevatron data. We compare our prediction to 
the more recent CDF data from run II in Refs.~\cite{Acosta:2004yw} (open
squares) and \cite{Abulencia:2006ps} (solid squares). In this figure the 
solid line presents the central prediction for 
$\xi_R=\xi_F=1$ and the dashed lines 
indicate the maximum and minimum values obtained by independently varying 
$\xi_R$ and $\xi_F=\xi_F^\prime$ in the range $1/2 \leq \xi_R,\xi_F \leq 2$ 
with the constraint $1/2 \leq \xi_R/\xi_F \leq 2$. The maximum and minimum 
values correspond to $\xi_F=2$ and $\xi_F=1/2$, respectively. The variation 
with $\xi_R$ is milder than the one with $\xi_F$. For $\xi_F < 1$, $\mu_F$ 
reaches the starting scale $\mu_0=m$ for the DGLAP evolution of the FFs and 
the $b$-quark PDF at $p_T=m_b\sqrt{1/\xi_F^2-1}$. For smaller values of $p_T$, 
there is no prediction because the FFs and the $b$-quark PDF are put to zero 
for $\mu_F < \mu_0$. This explains why the $p_T$ distribution for $\xi_F=1/2$ 
only starts at $p_T = \sqrt{3}m_b \approx 7.8$ GeV. The most recent data 
\protect\cite{Abulencia:2006ps} nicely agree with the GM-VFNS result. They lie 
close to the central prediction, with a tendency to fall below it in the lower
$p_T$ range, and they are comfortably contained within the theoretical error
band. We conclude from this, that the notorious Tevatron $B$-meson anomaly with
data-to-theory ratios of typically 2-3, that has been in the literature for
more than a decade, is actually not present thanks to both experimental and
theoretical progress. The previous CDF data \cite{Acosta:2004yw} based on the
measurement of $J/\psi +X$ final states are compatible with the latest ones for
$p_T < 12$ GeV, but are systematically below them for the larger values of 
$p_T$. This inconsistency becomes even more apparent by noticing that Fig. 4
only contains 4 out of the 13 data points for $p_T > 12$ GeV quoted in Ref.
\cite{Acosta:2004yw} and that the omitted data points line up with the selected
ones. This suggests that the systematical errors in Ref. \cite{Acosta:2004yw} 
and perhaps also in ref. \cite{Abulencia:2006ps}, might be underestimated and 
that the overall normalisation might need some adjustment.

The measured $p_T$ distributions of Ref. \cite{Acosta:2004yw} reaches down to
almost $p_T=0$ and exhibits a maximum at $p_T \approx 2.5$ GeV. This 
small-$p_T$ behaviour is correctly reproduced in the FFNS without DGLAP-evolved
FFs, which receive only contributions of class (i) without any subtractions.
It is clear that our present implementation of the GM-VFNS is not suitable
for cross sections in the small-$p_T$ region. Although the GM-VFNS is designed
to approach the FFNS in its region of validity without introducing additional
matching factors, to implement this numerically is not easy due to necessary
cancellations between different terms in the calculation. The problem to 
achieve such cancellations is complicated by the extra factorisation scale;
to obtain a smooth transition from the GM-VFNS to the FFNS, one has to 
carefully match terms that are taken into account at fixed order with terms 
that are resummed to higher orders in the PDFs and FFs.
In addition, it remains to be investigated whether a proper scale choice in
the small-$p_T$ range is required and helpful to ensure that the FFs and
$b$-quark PDF are sufficiently suppressed already at $p_T={\cal O}(m)$.

\begin{figure}[!thb]
\begin{center}
\epsfig{file=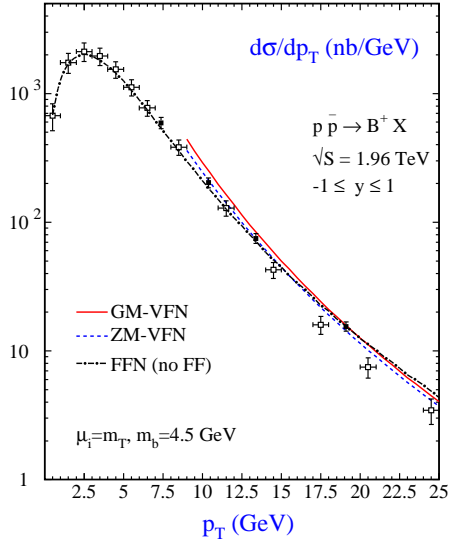,width=7.0cm}
\caption{\small Transverse-momentum distribution $d\sigma/dp_T$ of
$p\bar{p} \rightarrow B +X$ at c.m.\ energy $\sqrt{S}=1.96$~TeV integrated
over the rapidity range $|y| < 1$.
The central NLO predictions in the FFNS with $n_f=4$ and without FFs
(dot-dashed line), the ZM-VFNS (dashed line), and the GM-VFNS (solid line) are
compared with CDF data from Refs.~\cite{Acosta:2004yw} (open squares) and 
\cite{Abulencia:2006ps} (solid squares).}
\label{fig:fig6}
\end{center}
\end{figure}
We extend our numerical analysis to include the NLO prediction in the
FFNS with $n_f=4$ massless quark flavours in the initial state, which allows
us to also compare with the small-$p_T$ data from Ref.~\cite{Acosta:2004yw}.
We evaluate $\alpha_s^{(n_f)}(\mu_R)$ with $n_f=4$ and
$\Lambda^{(4)}=326$~MeV \cite{Pumplin:2002vw}, while we continue using the 
CTEQ6.1M proton PDFs \cite{Pumplin:2002vw}, in want of a rigorous FFNS set 
with $n_f=4$. In the FFNS, there is no room for DGLAP-evolved FFs, and only
$b,\bar b\to B$ transitions are included.
For simplicity, we identify $b$ (anti)quarks with $B$ mesons and account for
non-perturbative effects by including the branching fraction
$B(b\to B)=39.8\%$ \cite{pdg} as an overall normalisation factor, {\it i.e.}\
we use a $b\to B$ FF of the form $D(x)=B(b\to B)\delta(1-x)$, while the
$g,q,\bar q\to B$ FFs are put to zero.
In Fig.~\ref{fig:fig6}, the central FFNS (dot-dashed line), ZM-VFNS 
(dashed line), and GM-VFNS (solid line) predictions, for $\xi_R=\xi_F=1$, are 
compared with the CDF data from Refs.~\cite{Acosta:2004yw, Abulencia:2006ps}.
As in Fig.~\ref{fig:fig4}, some of the data points with $p_T>7$~GeV from
Ref.~\cite{Acosta:2004yw} are omitted for clarity.
Since the ZM-VFNS and our present implementation of the GM-VFNS are not
applicable to the small-$p_T$ range, we show the respective predictions only
for $p_T>2m=9$~GeV.
The GM-VFNS prediction shown in Fig.~\ref{fig:fig6} is identical with the 
central one in Fig.~\ref{fig:fig5}.
By construction, it merges with the ZM-VFNS prediction with increasing value
of $p_T$.
In accordance with the expectation expressed in the discussion of
Fig.~\ref{fig:fig4}, the difference between the GM-VFNS and
ZM-VFNS results is rather modest also at $p_T \gtrsim 2m$, since the 
$m$-dependent contribution, of class~(i), is numerically small and overwhelmed 
by the $m$-independent ones, of classes (ii) and (iii).
The FFNS prediction faithfully describes the peak structure exhibited by the
next-to-latest CDF data \cite{Acosta:2004yw} in the small-$p_T$ range and it 
also nicely agrees with the latest CDF data \cite{Abulencia:2006ps} way out to 
the largest $p_T$ values.
In fact, for $p_T>4m$, where its perturbative stability is jeopardised by
unresummed logarithms of the form $\ln(m_T^2/m^2) \gtrsim 3$, the FFNS 
prediction almost coincides with the GM-VFNS one, where such large logarithms 
are resummed.
This might be a pure coincidence, which becomes even more apparent if we also 
recall that the implementation of the $b,\bar b\to B$ transition in the FFNS is
not based on a factorisation theorem and quite inappropriate for such large 
values of $p_T$.
%
\section{Photoproduction}
Inclusive photoproduction of $D^\star$ mesons, $\gamma + p \to D^\star + X$, 
has been studied in Ref.\ \cite{Kramer:2003jw} where the direct part 
has been computed in the GM-VFNS whereas the resolved part has been 
included in the ZM-VFNS.
In this analysis the FFs of Ref.\ \cite{Binnewies:1998xq} and, for the 
resolved contribution, the GRV92 photon PDFs 
\cite{cit:GRV-9202} have been utilized. The other parameters have been chosen 
as specified in Sec.\ \ref{sec:parameters}.
In Fig.\ 6 of Ref.\ \cite{Kramer:2003jw}, the central numerical
predictions for the $p_T$ distributions of the $D^\star$ meson have been 
compared with preliminary ZEUS data \cite{Zeus:photo}.
There exist similar data by the H1 collaboration
\cite{H1:photo} which have not been used in this analysis.
As can be seen in this figure, the agreement of the 
$p_T$-distributions with the data is quite good down to
$p_T \simeq 2 m_c$ and the mass effects turn out to be small.
In order to extend the range of applicability of the GM-VFNS
into the region $p_T < 3\ \gev$ more work on the matching to
the 3-fixed flavour theory would be needed.
Figs.\ 7 -- 9 of  Ref.\ \cite{Kramer:2003jw},
showing results for the rapidity ($y$), invariant mass ($W$) and
inelasticity ($z(D^\star)$) distributions, have to be taken with a
grain of salt since they receive large contributions from the 
transverse momentum region $1.9 < p_T < 3\ \gev$ which is outside 
the range of validity of the present theory.
\begin{figure*}[t]
\begin{center}
\begin{tabular}{ll}
{\parbox{5.0cm}{
\hspace*{-1.2cm}
\epsfig{file=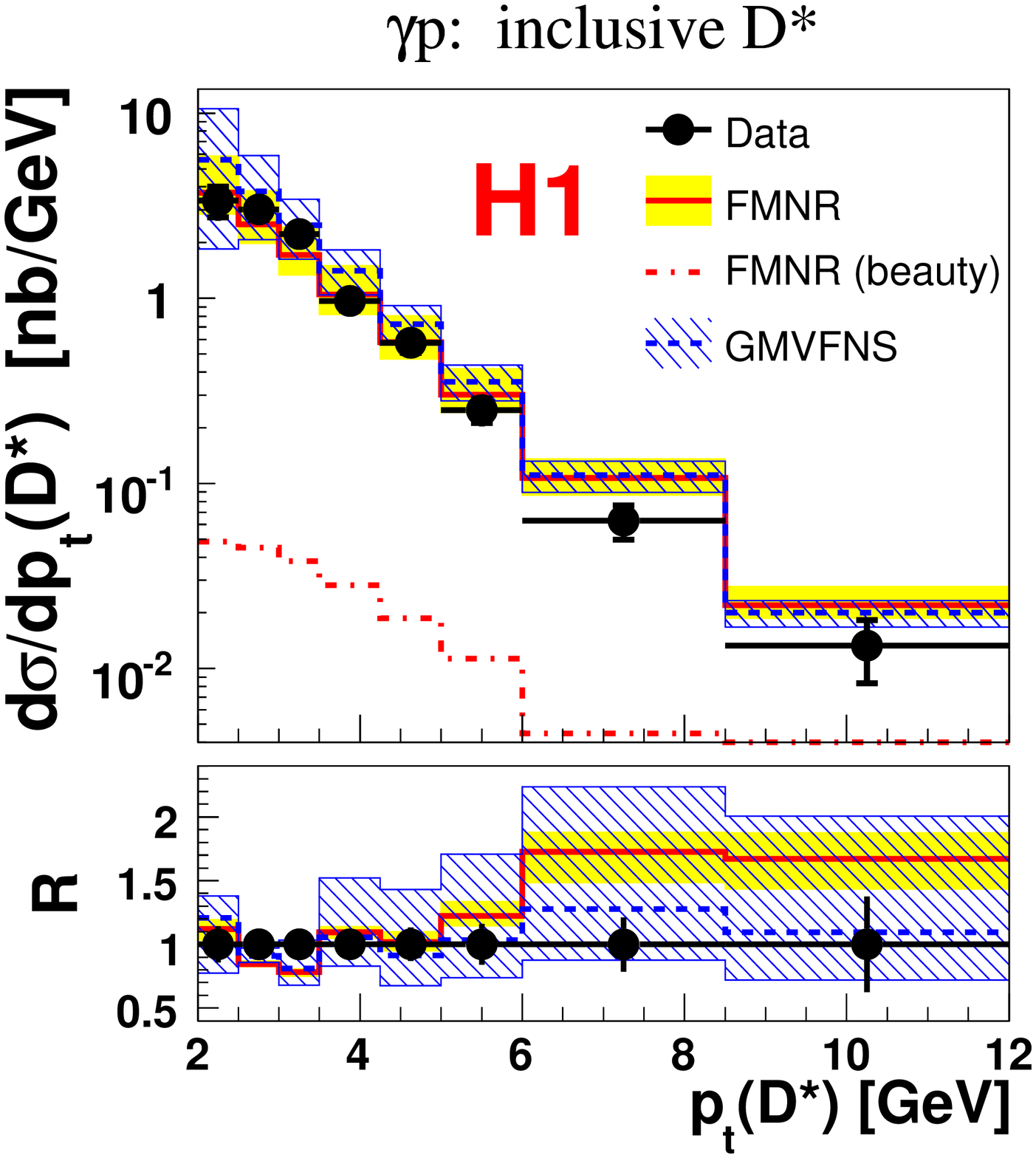,width=6.0cm}
}} &
{\parbox{5.0cm}{
\epsfig{file=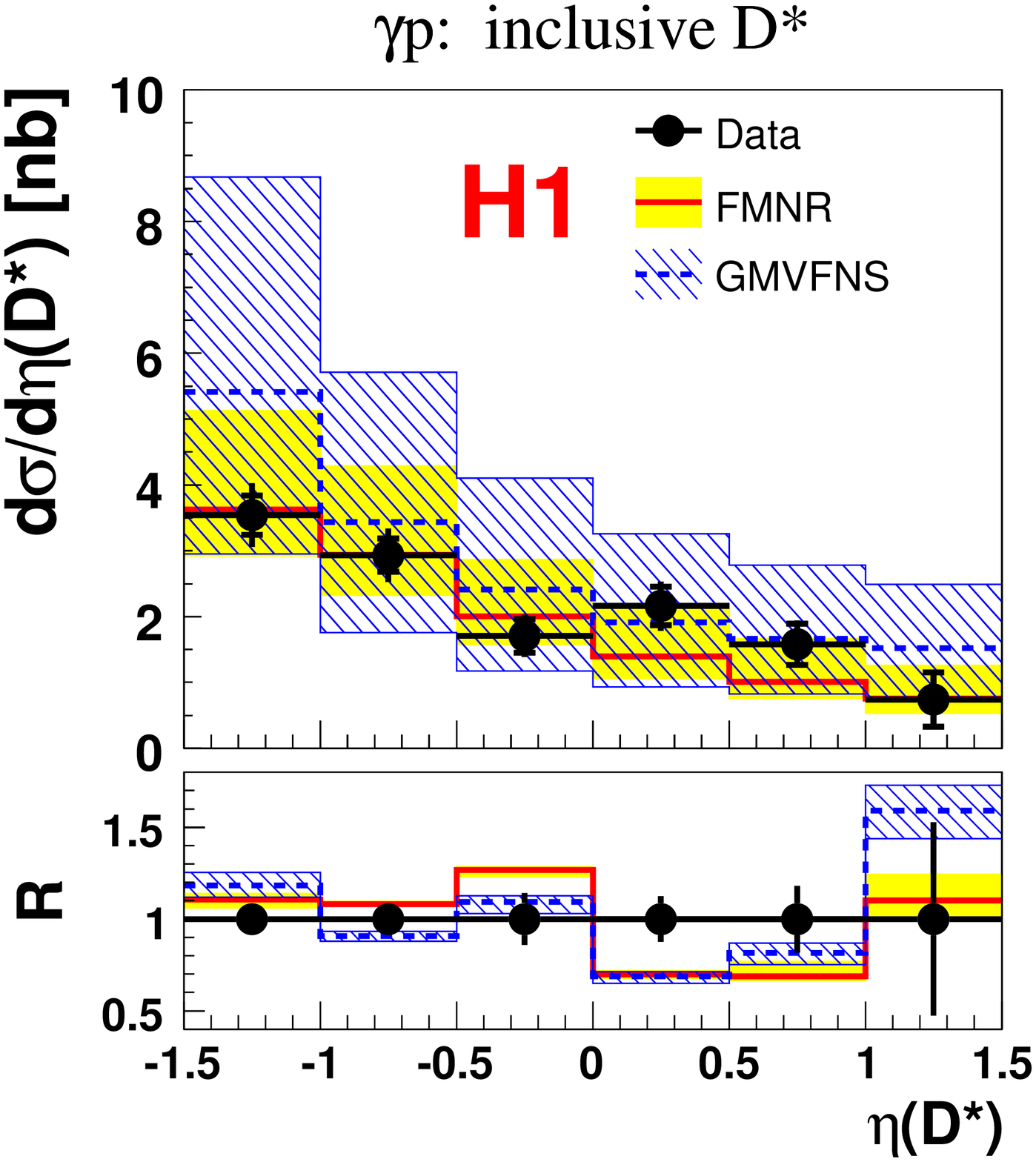,width=6.0cm}
}} 
\end{tabular}
\end{center}
\caption{
Inclusive $D^*$ cross sections as a function of $p_t(D^*)$ (left) and 
$\eta(D^*)$ (right) compared to NLO QCD calculations of FMNR 
\protect\cite{Frixione:1995qc} in the FFNS and GM-VFNS for 
photoproduction in the laboratory frame. The FMNR bottom contribution is 
shown separately for the $p_t(D^*)$ distribution.
}
\label{fig:fig7}
\end{figure*}
With the work in Ref.\ \cite{Kniehl:2004fy},
it was possible to include also the resolved part in the GM-VFNS.
This has been done and the predictions in the complete GM-VFNS framework at 
NLO, combined with updated FFs \cite{Kniehl:2006mw}, have been compared with
recent H1 photoproduction data \cite{Aktas:2006ry}. The results of the 
calculation and the comparison with the data is shown in Fig.~\ref{fig:fig7}. 
In this figure also the predictions in the FFNS based on the FMNR program
\cite{Frixione:1995qc} are shown. The experimental cross section as a function
of $p_T$ falls steeply with increasing $p_T$ as predicted by
both calculations. FMNR predicts a distribution which decreases less steeply
at large $p_T$ than the data as is seen more clearly from the plot of the ratio
of the theoretical over the measured cross section. Also in Fig.~\ref{fig:fig7}
the differential cross section as a function of the pseudorapidity 
$\eta(D^*)$ is shown. This cross section decreases with increasing $\eta$. 
Both calculations predict a similar shape and agree nicely with the data. The 
GM-VFNS prediction shows a larger scale dependence. Otherwise the two 
calculations give rather similar results, which is remarkable, considering the 
very different ingredients of the two approaches.  

\section{Summary}
We have discussed one-particle inclusive production of heavy-flavoured
hadrons in hadron--hadron and photon--proton collisions
in a massive variable-flavour-number scheme (GM-VFNS).
The importance of a unified treatment of all these processes, based on
QCD factorisation theorems, has been emphasised, in order to provide
meaningful tests of the universality of the FFs and hence of QCD.
At the same time, it is necessary to incorporate heavy-quark mass effects
in the formalism since many of the present experimental data points lie in
a kinematical region where the hard scale of the process is not much
larger than the heavy-quark mass.
This is achieved in the GM-VFNS, which includes heavy-quark mass effects
and still relies on QCD factorization.
We have discussed numerical results for two reactions.
In general, the description of the transverse momentum spectra is
quite good down to transverse momenta $p_T \simeq 2 m$.
Extending the range of applicability of our scheme to smaller
$p_T$ would require more work on the matching to the corresponding
theories in the fixed flavor number scheme.

\section*{Acknowledgments}
The author would like to thank the organisers 
of the DIS 2007 workshop 
for the kind invitation, 
B.\ A.\ Kniehl, I.\ Schienbein and H.\ Spiesberger 
for their collaboration.

\begin{footnotesize}

\end{footnotesize}
\end{document}